\documentclass[fleqn,10pt]{wlscirep}
\usepackage[utf8]{inputenc}
\usepackage[T1]{fontenc}

\title{An OpenMetBuoy dataset of Marginal Ice Zone dynamics collected around Svalbard in 2022 and 2023}

\author[1,*]{Jean Rabault}
\author[2]{Catherine Taelman}
\author[3]{Martina Idžanović}
\author[4]{Gaute Hope}
\author[5]{Takehiko Nose}
\author[6]{Yngve Kristoffersen}
\author[7]{Atle Jensen}
\author[4,8]{Øyvind Breivik}
\author[4,8]{Helge Thomas Bryhni}
\author[9]{Mario Hoppmann}
\author[10]{Denis Demchev}
\author[11]{Anton Korosov}
\author[2]{Malin Johansson}
\author[2]{Torbjørn Eltoft}
\author[4]{Knut-Frode Dagestad}
\author[3]{Johannes Röhrs}
\author[10]{Leif Eriksson}
\author[12]{Marina Durán Moro}
\author[3,13]{Edel S. U. Rikardsen}
\author[5,14]{Takuji Waseda}
\author[5]{Tsubasa Kodaira}
\author[2]{Johannes Lohse}
\author[15]{Thibault Desjonquères}
\author[2]{Sveinung Olsen}
\author[7]{Olav Gundersen}
\author[2,3]{Victor Cesar Martins de Aguiar}
\author[2]{Truls Karlsen}
\author[16]{Alexander Babanin}
\author[16]{Joey Voermans}
\author[17]{Jeong-Won Park}
\author[13,18]{Malte Müller}

\affil[1]{Norwegian Meteorological Institute, IT Department, Oslo, 0313, Norway}
\affil[2]{UiT The Arctic University of Norway, Dept. of Physics and Technology, 9037 Tromsø, Norway}
\affil[3]{Norwegian Meteorological Institute, Division for Ocean and Ice, 0371 Oslo, Norway}
\affil[4]{Norwegian Meteorological Institute, Division for Oceanography and Marine Meteorology, 5007 Bergen, Norway}
\affil[5]{The University of Tokyo, Graduate School of Frontier Sciences, Kashiwa, 277-8561, Japan}
\affil[6]{University of Bergen, Department of Earth Science, Bergen, 5007, Norway}
\affil[7]{University of Oslo, Department of Mathematics, Oslo, 0313, Norway}
\affil[8]{University of Bergen, Geophysical Institute, 5007 Bergen, Norway}
\affil[9]{Alfred-Wegener-Institut, Helmholtz-Zentrum fuer Polar- und Meeresforschung, 27580 Bremerhaven, Germany}
\affil[10]{Department of Space, Earth and Environment, Chalmers University of Technology, 41296 Gothenburg, Sweden}
\affil[11]{Ocean and Sea Ice Remote Sensing Group, Nansen Environmental and Remote Sensing Center, 5006 Bergen, Norway }
\affil[12]{Norwegian Meteorological Institute, Remote sensing and data management division, Oslo, Norway}
\affil[13]{University of Oslo, Department of Geosciences, Oslo, 0313, Norway}
\affil[14]{Japan Agency for Marine-Earth Science and Technology, Yokosuka, 237-0061, Japan}
\affil[15]{Gothenburg University, 40508 Gothenburg, Sweden}
\affil[16]{Department of Infrastructure Engineering, University of Melbourne, Melbourne, Australia}
\affil[17]{Center of Remote Sensing and GIS, Korea Polar Research Institute, Incheon, 21990, South Korea }
\affil[18]{Norwegian Meteorological institute, R\&D Department, Oslo, 0313, Norway.}

\affil[*]{corresponding author: Jean Rabault (jean.rblt@proton.me)}

\begin{abstract}
Sea ice is a key element of the global Earth system, with a major impact on global climate and regional weather. Unfortunately, accurate sea ice modeling is challenging due to the diversity and complexity of underlying physics happening there, and a relative lack of ground truth observations. This is especially true for the Marginal Ice Zone (MIZ), which is the area where sea ice is affected by incoming ocean waves. Waves contribute to making the area dynamic, and due to the low survival time of the buoys deployed there, the MIZ is challenging to monitor. In 2022-2023, we released 79 OpenMetBuoys (OMBs) around Svalbard, both in the MIZ and the ocean immediately outside of it. OMBs are affordable enough to be deployed in large number, and gather information about drift (GPS position) and waves (1-dimensional elevation spectrum). This provides data focusing on the area around Svalbard with unprecedented spatial and temporal resolution. We expect that this will allow to perform validation and calibration of ice models and remote sensing algorithms.

\end{abstract}
\begin{document}

\flushbottom
\maketitle

\thispagestyle{empty}

\section*{Background \& Summary}

Sea ice is an important component of the global climate system. Over 7\% of the global ocean surface is, on average over a year, covered by sea ice \cite{parkinson_earth_1997}. Sea ice influences several key couplings between the ocean and the atmosphere, including heat and momentum transfers \cite{zhang_increasing_2007,tersigni_high-resolution_2023,muilwijk_future_20242024-09-Isfjorden-Mooring-01}. Sea ice is also a key element of several self-amplifying feedback couplings: for example, the albedo of the polar oceans is strongly dependent on the presence of sea ice, which also controls the amount of solar energy absorbed by the ocean \cite{shao_springsummer_2015}. Similarly, sea ice controls many biological phenomena, e.g. by modulating the amount of solar energy available in the upper water column \cite{duncan_spatio-temporal_2024}, and providing an habitat for many species \cite{laidre_arctic_2015,arrigo_sea_2017}.

Sea ice dynamics are challenging to model and predict. This is due both to the complexity of the underlying physics \cite{roach2024physics,squire_fresh_2018}, which involves a broad range of phenomena including sea ice formation \cite{weeks_growth_1986,dieckmann2003importance}, melting \cite{markus2009recent,wadhams2004ocean}, drift \cite{spreen2011trends,sutherland_estimating_2022}, fracture and breakup \cite{kohout_storm-induced_2014,voermans_experimental_2020}, and to the logistical and practical difficulties involved in studying sea ice, which limit the amount of in-situ data available \cite{rabault_dataset_2023}. This second point is especially true in the Marginal Ice Zone (MIZ), i.e., the sea ice area strongly affected by open water phenomena, in particular waves \cite{squire_ocean_2020}. Indeed, buoys deployed there are subject to violent dynamics such as ice breakup, ridging, and crushing, and have, therefore, a statistically limited life expectancy.

The limited amount of in-situ data in the MIZ is in stark contrast to the importance and richness of the wave-ice interaction mechanisms \cite{squire_ocean_2020,roach2024physics}. The combination of wave diffraction \cite{squire_ocean_2020}, wave reflections \cite{fox_reflection_1990}, visco-elastic dissipation in the ice \cite{zhao_wave_2015,cheng_calibrating_2017}, turbulent attenuation under the ice \cite{herman_spectral_2021,voermans_wave_2019,rabault_experiments_2019}, sea ice breakup \cite{kohout_storm-induced_2014,voermans_experimental_2020}, and ice-ice collisions \cite{herman_wave-induced_2018,loken_experiments_2022,dreyer_direct_2024} creates complex dynamics that require significant amounts of data to be disentangled. Arguably, the spatial and temporal resolution of existing datasets is still too low to obtain a statistically well-sampled representative overview of the sea ice dynamics in the MIZ. This, in turn, has traditionally limited the quality of e.g. operational models used to describe and predict MIZ dynamics \cite{squire_fresh_2018,nose_comparison_2023}. Such predictions are critical to enable better weather and climate models, as well as safe and environment-friendly human activities in ice-infested and ice-covered regions, which is a key objective in a context of global warming \cite{wunderling_global_2020} and a strong increase in human activity levels in the Arctic \cite{muller_arctic_2023}.

Consequently, considerable effort has been devoted over several decades to collecting increasing amounts of in-situ data from the MIZ \cite{doble_wave_2006,kohout_device_2015,rabault_dataset_2023}. In this regard, a critical aspect lies in being able to deploy enough ice buoys to compensate for the limited life expectancy of each individual instrument, and so to still manage to collect datasets large enough to provide a good statistical overview of the MIZ \cite{thomson_overview_2018,rogers_estimates_2021}. In recent years, considerable technical progress has been made, which has allowed to cut the cost of individual buoys by over an order of magnitude through the use of off-the-shelf components and open source designs \cite{kohout_device_2015,rabault_open_2020,rabault_openmetbuoy-v2021_2022,thomson_development_2024}. This, in turn, has a significant impact on the accuracy of state-of-the-art wave in ice parameterizations \cite{rogers_estimates_2021}, and is resulting in a drastic improvement in operational model skills in the Arctic MIZ region \cite{aouf_validation_2024}. However, despite these recent advances, there is still room for 
considerable improvement \cite{nose_comparison_2023}.

\begin{figure}[ht]
\centering
\includegraphics[width=\linewidth]{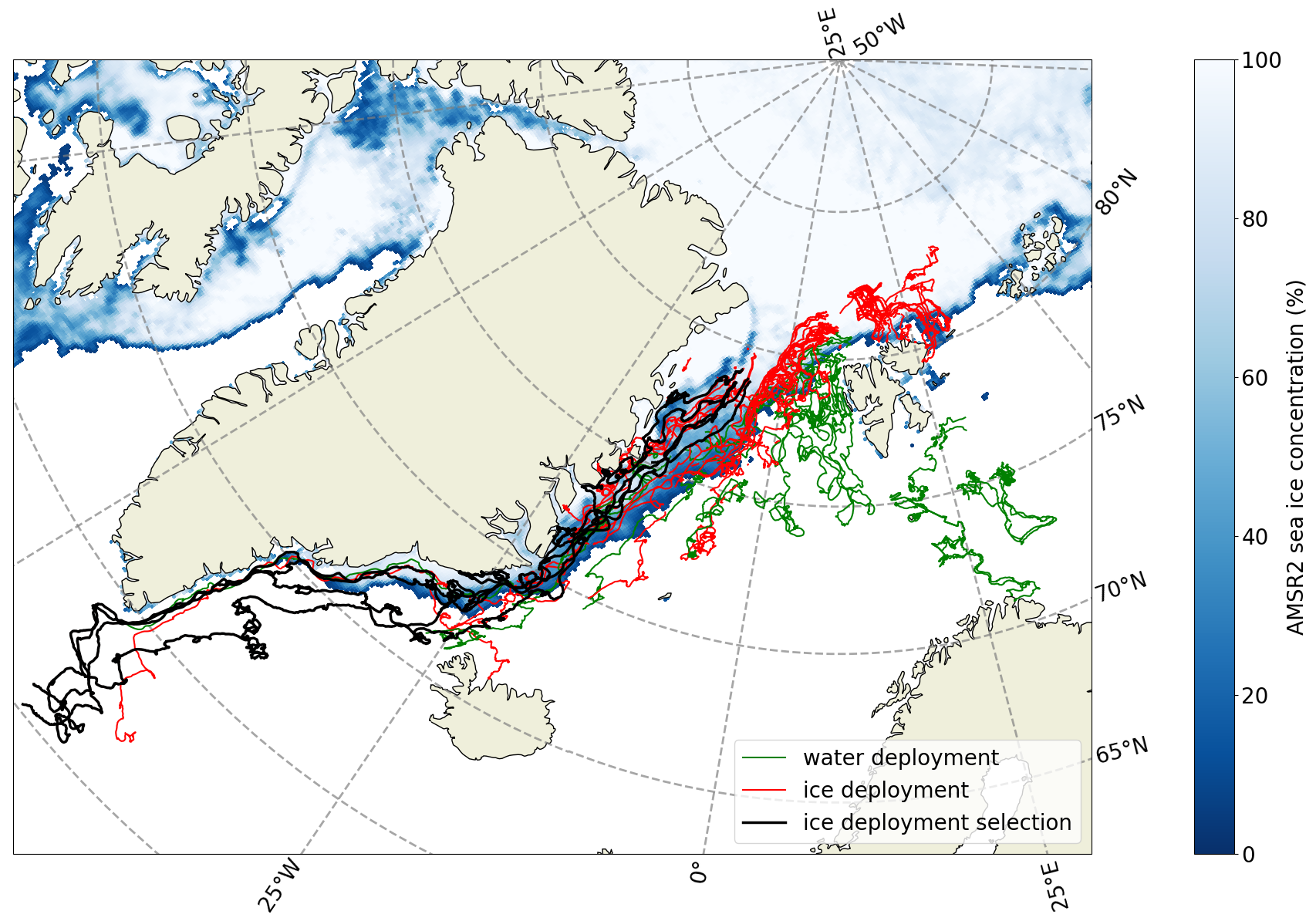}
\caption{Summary of the 80 trajectories released in the present dataset. Buoys that were initially deployed on the sea ice correspond to the red trajectories, while buoys that were initially deployed in the open water outside of the sea ice correspond to the green trajectories. In addition, a subset of 5 buoys initially deployed on the sea ice, which spectra are shown in Fig. \ref{fig:overview_spectra}, correspond to the black trajectories. The sea ice concentration obtained from the AMSR2 dataset on June 15th 2022 is plotted as background.}
\label{fig:overview_trajectory}
\end{figure}

\begin{figure}[ht]
\centering
\includegraphics[width=\linewidth]{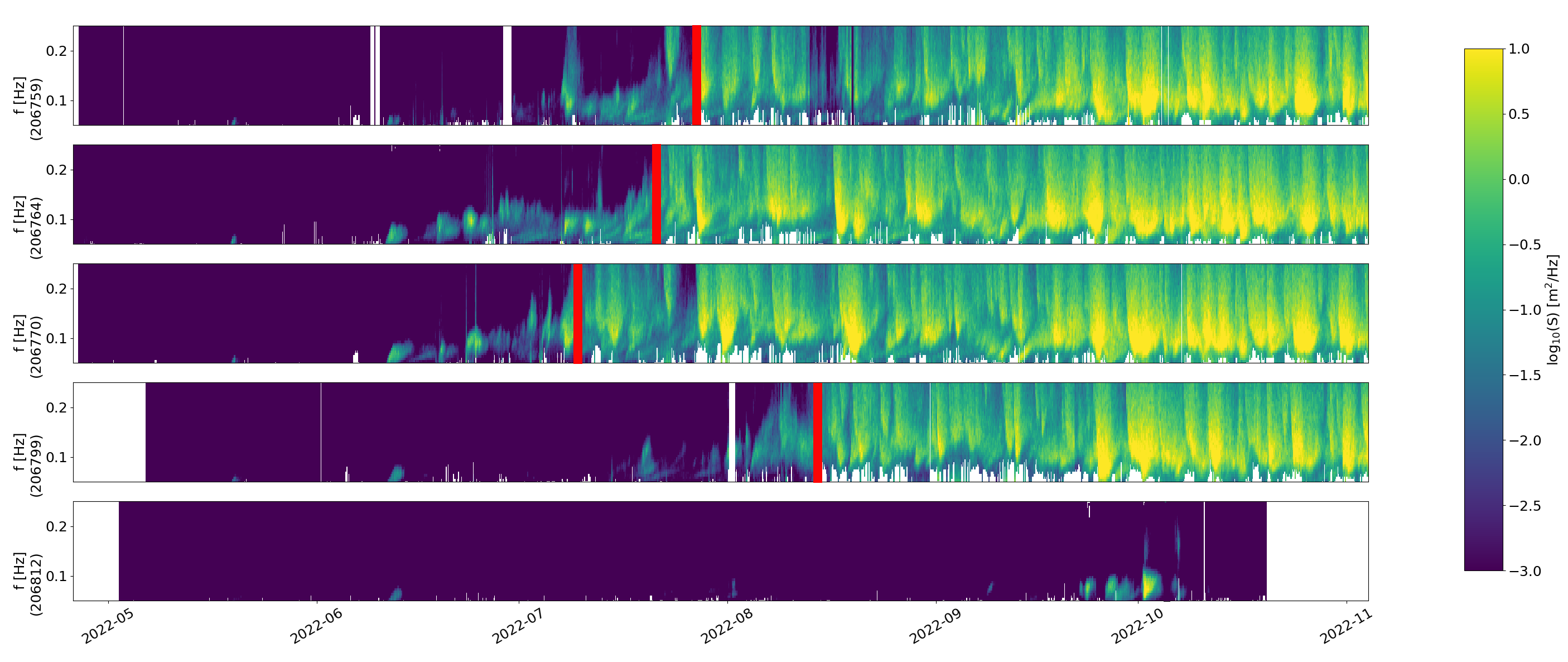}
\caption{Illustration of the wave spectra acquired by a subset of 5 buoys from the CIRFA\_UIT deployment. The buoys were initially deployed on sea ice in late April to early May 2022. The 4 buoys at the top ultimately ended up in water, as visible by the transitions (highlighted by the thick red bars) from spectra where high frequencies are filtered out, to spectra where significant high frequency energy is present. The 5th buoy never ended up in the open water, as it drifted into an ice-covered fjord on the East coast of Greenland. Transmission took place until mid December 2022 for the 4 buoys that survived transition to open water conditions.}
\label{fig:overview_spectra}
\end{figure}

In the present dataset, we release 79 trajectories collected by OpenMetBuoys (OMBs, \cite{rabault_openmetbuoy-v2021_2022,rabault_tracking_2022}) in the region around Svalbard, over the years 2022-2023. This is, to our knowledge, the largest single and consistent dataset on drift and waves in a specific MIZ area to date. Our general dataset is illustrated in Fig. \ref{fig:overview_trajectory}. A sample of spectra collected by the buoys can be observed in Fig. \ref{fig:overview_spectra}. We expect that these data will be critical to enable the tuning, calibration, and validation of a new generation of sea ice parameterizations and models, unlocking further development and improvement of operational weather forecasts and climate models in the Arctic. Moreover, these data will allow to calibrate and validate algorithms used for remote sensing of sea ice drift and waves in ice in the MIZ, which is a very active area following the launch of several state-of-the-art satellite systems \cite{collard_wind-wave_2022,peureux_sea-ice_2022}.


\section*{Methods}

The present dataset consists entirely of OpenMetBuoys-v2021. The OMB is a free and open source software and hardware (FOSSH) instrument platform developed to collect oceanographic data and refined for specific applications by a growing international community of scientists and makers \cite{rabault_openmetbuoy-v2021_2022}. The OMB and similar working buoys have been discussed in several previous studies, including \cite{rabault_dataset_2023,kodaira_development_2022}, and the key sensor and algorithmic techniques used have been refined and established over many years, see e.g. \cite{kohout_device_2015,thomson_development_2024,rabault_open_2020,sutherland_observations_2016,rabault_measurements_2016,rabault_measurements_2017}. As a consequence, the OMB design and core features have been extensively described and validated on many occasions, and we only provide a general overview of the OMB characteristics and functions in the following to avoid repeating boilerplate technical information and validation data. The reader curious about the electronics and algorithmic details is referred to the technical paper introducing and validating the OMB \cite{rabault_openmetbuoy-v2021_2022}, as well as to the open source code on GitHub (see the links in Table \ref{tab:urls}), and previous data released using the same OMB design \cite{rabault_dataset_2023}. The OMB and related platforms have been used in several scientific studies presented by a range of groups; see, e.g., \cite{voermans_wave_2021,nose_comparison_2023,nose_observation_2024,rabault_-situ_2024,dreyer_direct_2024}. To date, the authors have been informed about the deployment of over 400 individual OMBs by 12 different groups around the world, though even more OMB-based or OMB-inspired buoys may have been built and deployed without the authors' knowledge, owing to the fully free and open nature of the design.

The OMB is self-contained in a small box measuring around 10 $\times$ 10 $\times$ 12 cm, and weighting between 0.5 and 1 kg depending on the amount of batteries included. The design is energy efficient enough that it can be operated on non-rechargeable lithium batteries for extended periods of time without the need for a solar panel. With 3 D-cell lithium batteries and standard sampling rates, a typical lifetime of up to over 7.5 months is observed under low-temperature polar conditions. By including more batteries, deployment durations of up to 1 year have been obtained in stable Antarctic ice \cite{nose_observation_2024}. Deployments taking place around Svalbard are typically more limited in duration, owing to the rapid drift speed of the sea ice towards the open ocean, which is characteristic for this area. The OMB uses iridium short burst data (SBD) satellite communications to enable global communications.

\begin{figure}[ht]
\centering
\includegraphics[width=0.49\linewidth]{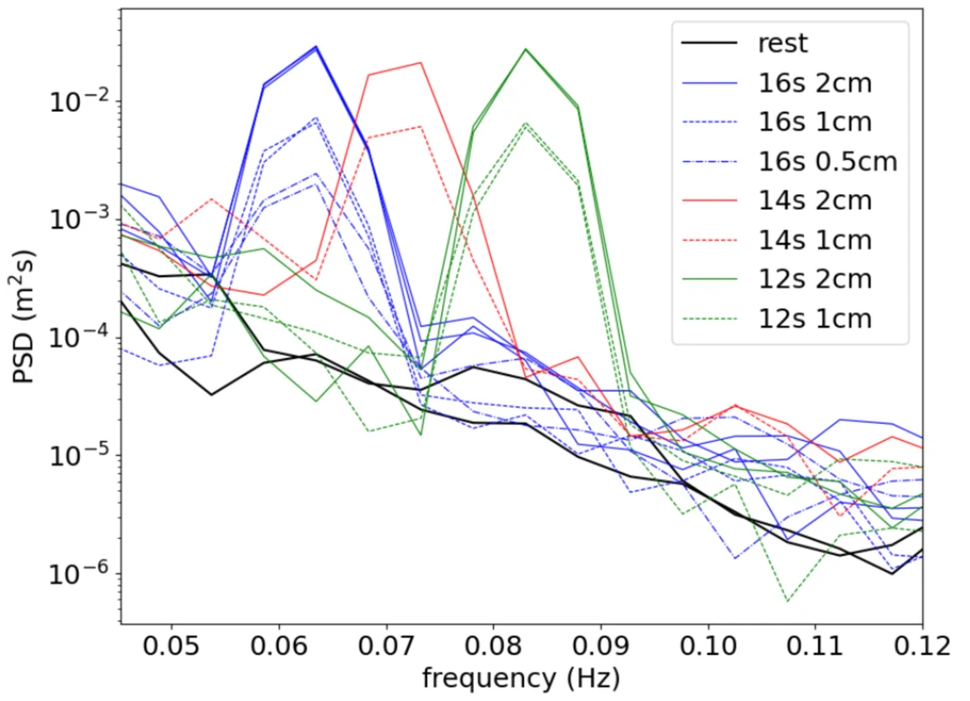}
\caption{Noise threshold, both at rest and under wave conditions, in a test experiment in the laboratory. This confirms the ability of the OMB-v2021 to reliably measure waves down to typically 0.5 cm amplitude and 16 s period. Reproduced from \cite{rabault_openmetbuoy-v2021_2022}.}
\label{fig:wave_accuracy}
\end{figure}

The core OMB functionality is enabled by combining off-the-shelf, consumer grade microelectromechanical system (MEMS) sensors, and open source, custom on-board processing. The brain of the OMB is a single microcontroller unit (MCU, Ambiq Apollo 3) that controls all sensors and performs real-time signal processing. Drift and trajectory data are obtained from a multiband Global Navigation Satellite System (GNSS) module (Zoe-M8Q). The typical position accuracy indicated by the datasheet is around $\pm 2.5$ m to $\pm 4$ m, depending on the satellite constellation available, which corresponds to our experience in the field. Wave measurements are performed by a MEMS motion sensor (ISM330DHCX from STMicroelectronics), that performs both acceleration and angular rate measurements at a frequency of 800 Hz. The MCU performs on-the-fly Kalman filtering, 2-stage lowpass filtering, and quaternion arithmetics, so that the filtered vertical buoy acceleration (in the Earth frame of reference) is obtained at 10 Hz. A time series of 12288 consecutive vertical acceleration measurements (to be a multiple of 2048 and allow fast Fourier transform), corresponding to around 20 minutes of vertical acceleration data, is used to generate each 1-dimensional (1D) vertical acceleration spectrum using the Welch method. The resulting spectrum can then either be used as is or converted to a 1D vertical elevation spectrum (see Equation 1 in \cite{rabault_measurements_2016}). The aggregated system accuracy allows us to measure waves down to around 0.5 cm amplitude and 16 seconds period with a signal-to-noise ratio of over 10, as proved in \cite{rabault_openmetbuoy-v2021_2022} (see Fig. \ref{fig:wave_accuracy} reproduced from there), and as observed experimentally in the field, e.g., \cite{nose_observation_2024}. This allows us to observe attenuated swells up to over 1000 km into the ice in stormy conditions in the Antarctic ocean. The key properties of the OMB are summarized in Table \ref{tab:omb_properties}.

The exclusive use of off-the-shelf components and open source software and hardware allows to produce an OMB for typically under 650 USD hardware cost, with a total cost of ownership (TCO) of typically around 1000 USD when including three months of operation in the Arctic (in particular the corresponding Iridium communication costs). Building instructions are available at \url{https://github.com/jerabaul29/OpenMetBuoy-v2021a} (all URLs are also gathered, for clarity, in Table \ref{tab:urls}). This is, to the best of our knowledge, around one order of magnitude less expensive than currently existing turn-key commercial alternatives. A commercial OMB version can also be bought as a turn-key product for slightly above 1000 USD (see the LabMaker Gmbh product description, \url{https://www.labmaker.org/products/openmetbuoy}; note that the authors have no financial links with the company). This cost reduction is key in achieving large scale deployments on tight research budgets, and is the core feature that enables us to release the present dataset of 79 trajectories.

\begin{figure}[ht]
\centering
\includegraphics[width=0.49\linewidth]{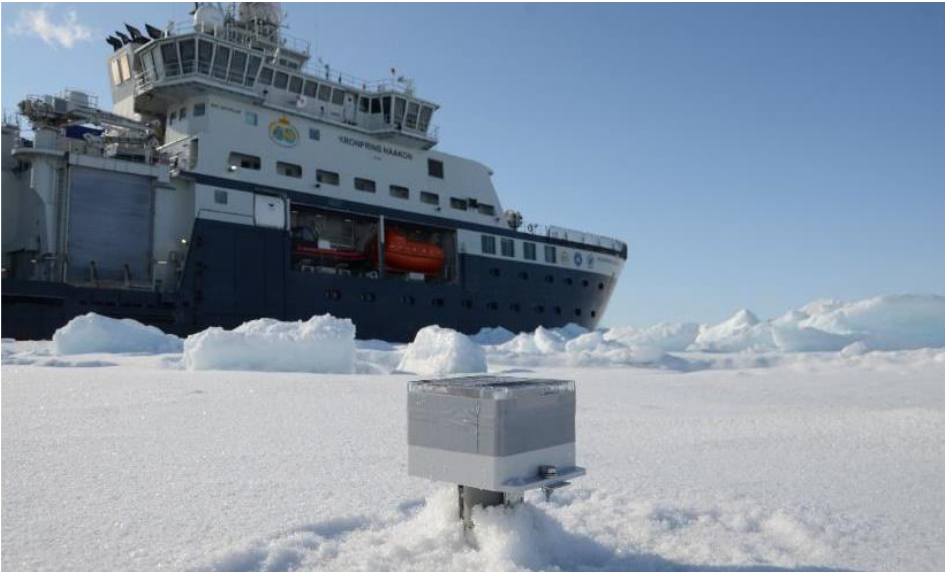}
\includegraphics[width=0.445\linewidth]{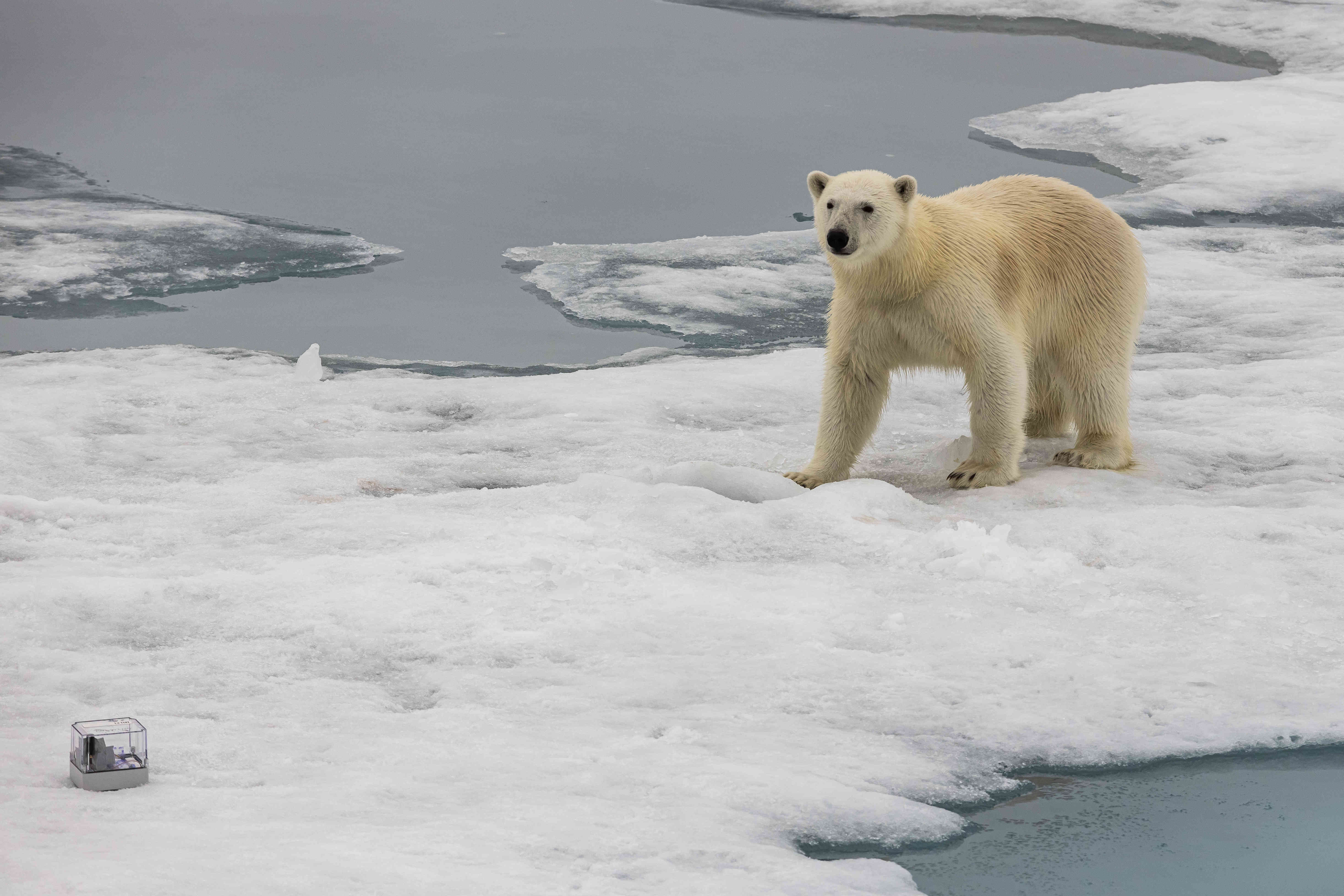}
\caption{Left: illustration of the OMB deployment on top of a frozen-in mount to elevate the box over the sea ice. Picture taken in the context of the CIRFA\_UIT deployment, credits Johannes Lohse, UiT, reproduced from \cite{wolfgang_dierking_2022_7314066}. Right: illustration of the OMB deployment directly on the sea ice. Picture taken in the context of the AWI\_UTOKYO deployment, credits Mario Hoppmann.}
\label{fig:example_deployment}
\end{figure}

OMBs are suitable for a variety of deployment cases. More specifically, OMBs can be deployed in the open ocean by being simply dropped into the water. With the standard housing and battery quantities, the OMB floats well enough to accurately measure waves, as its small footprint implies that its response amplitude operator (RAO, \cite{newman_marine_1977}) does not impact the hydrodynamic response in the wave frequency range measured \cite{rabault_openmetbuoy-v2021_2022}. When deployed on sea ice or icebergs, OMBs can be either put directly on top of the snow or ice layer (whichever is observed at any given location), or mounted on a support frozen into the ice if it should be elevated over the snow ice surface. The choice of one or the other solution is mostly a balance between the local snowfall expectation and the risk of attracting polar bear attention (see Fig. \ref{fig:example_deployment}). In regions with heavy snowfalls, putting the OMB on an elevated support is necessary to mitigate the risk of the buoy being buried under snow and loosing the ability to communicate over satellite, but a higher profile means more risk of attracting (destructive) polar bear attention.

%

\section*{Data Records}

The data for each deployment included in the present dataset are made available as netCDF-CF files on the Arctic Data Center hosted by the Norwegian Meteorological Institute at the following URL: \url{https://adc.met.no/datasets/10.21343/w2se-b681}. Moreover, all the raw binary Iridium SBD message data, processing scripts used to analyze these, Python scripts used to generate the plots of this article, and packaging scripts used to generate the netCDF files, are available on GitHub at the following URL: \url{https://github.com/jerabaul29/2024_OpenMetBuoy_data_release_MarginalIceZone_SeaIce_OpenOcean}. The data processing pipelines used to convert the binary Iridium SBD messages into netCDF-CF files are provided by the open source Trajectory Analysis (TrajAn) package and its OMB reader integration, which are available at: \url{https://github.com/OpenDrift/trajan}.

In the following, we present a brief overview of the different deployments included in this data release. The reader looking for the detailed cruise reports (when applicable, depending on deployment) is referred to the GitHub repository at \url{https://github.com/jerabaul29/2024_OpenMetBuoy_data_release_MarginalIceZone_SeaIce_OpenOcean}. The trajectory length is dependent from buoy to buoy, which is, to the best of our knowledge, the consequence of the harsh conditions in the dynamic Arctic MIZ (including snowfalls, polar bears destroying instrumentation, sea ice melting and breakup crushing and sinking buoys, heavy storm conditions, burial under heavy snowfall, etc.), rather than the consequence of technical issues per se.

The list of deployment names, in chronological order (corresponding to the start of deployments), is as follows:

\begin{itemize}
    \item CIRFA\_UIT: deployment of 19 OMBs on the ice in the context of the 2022 CIRFA cruise \cite{wolfgang_dierking_2022_7314066}. The instruments were deployed in April and May 2022 in the sea ice area in western Fram Strait. Sixteen OMBs were manually deployed on drifting sea ice floes using elevated mounts to avoid being buried by snowfalls. The three remaining OMBs were deployed on the top of large icebergs using drones and a remotely controlled hook release system to put them in position. Since these three OMBs had strict weight requirements, they carried less batteries and the wave measurement part was disabled to save power. Data transmissions took place until December 2022, though the instruments still transmitting had ended up in water by then. Auxiliary data, including in-situ snow and ice properties at the deployment sites and a list of overlapping Sentinel-1 SAR imagery for each OMB trajectory, are available at \cite{DDGNWA_2023}.
    \item CIRFA\_METNO: deployment of 15 OMBs in the open water outside of the MIZ in the context of the 2022 CIRFA cruise. These buoys were deployed in the open water immediately outside of the area where the CIRFA\_UIT ice buoys were deployed, to offer a possibility to compare drift between ice and open water conditions. The instruments were deployed in April 2022 in the open water area between Svalbard and Greenland. Seven OMBs were equipped with wave measurement capability (i.e., the ones having a \_ISM or \_LSM suffix in their name, see the released netcdf files), while the other ones only performed GNSS position measurements to save costs.
    \item AWI\_UTOKYO: deployment of 15 OMBs as a collaboration between the Alfred Wegener Institute and the University of Tokyo. The OMBs were deployed on the sea ice Northwest of Svalbard in July 2022 and transmitted until October 2022.
    \item UIO\_DOFI: deployment of 14 OMBs on the sea ice North of Svalbard from the Research Hovercraft R/H Sabvabaa. The buoys were deployed on the sea ice in August 2022, and transmissions took place until November 2022, though the buoys were either stranded or had ended up in open water by then.
    \item CAGE: deployment of six buoys in the open water around the Bjørnøya area in the context of the CAGE experiments. The deployment took place in August 2022 and transmissions lasted until November 2022, and more information and data are available at \cite{de_aguiar_drifter_2024}.
    \item UIB\_METNO: deployment of two OMBs in the water North of the Hopen island. The deployment took place in October 2022 and data transmission took place until December 2022.
    \item CHALMERS: deployment of seven OMBs on the sea ice in the area between Svalbard and Greenland. The deployment took place in May 2023, and one of the buoys continued transmitting until November 2023, though it had ended up in open water by then.
    \item AWI\_UOM: deployment of 1 OMB on sea ice about 80 km from the ice edge North of Svalbard as a collaboration between the Alfred-Wegener-Institute and the University of Melbourne. The OMB was deployed in August 2023 and last transmission occurred on 1 November 2023. Based on the wave spectra high frequency tail the buoy ended in open water at the end of October.
\end{itemize}

The deployments are summarized in Table \ref{tab:summary}. Moreover, all instrument trajectories, colored by the kind of deployment for each individual buoy (on sea ice or in water), are presented in Fig. \ref{fig:overview_trajectory}. The background color indicates the sea ice concentration as retrieved from the AMSR2 satellite product \cite{spreen_sea_2008,melsheimer_amsr-e_2020}, for 2022-06-15T12Z. Note that the sea ice concentration background is a snapshot at a given point in time, while the trajectories extend over a long period of time. Despite this fact, the sea ice concentration indicated is typically representative of the late spring sea ice extent in the area. Fig. \ref{fig:overview_spectra} presents an illustration of the typical wave data collected by the buoys, focusing on displaying a subset of the OMBs positioned on the ice during the CIRFA\_UIT deployment for clarity and ease of reading.

%

\section*{Technical Validation}

The OpenMetBuoy-v2021 used here has been validated and used in many studies before, see in particular the technical development paper \cite{rabault_openmetbuoy-v2021_2022}, the previous data release paper \cite{rabault_dataset_2023}, and a number of scientific publications \cite{nose_comparison_2023,nose_observation_2024}. The OMBs used here have exactly identical design from both a hardware and software perspective and, therefore, no more technical validation is necessary, and we refer the reader curious of technical details to the corresponding publications. The GNSS position measurement is provided directly by a well-established chip, and no more technical validation is needed. The wave vertical spectrum and statistics are computed using an established methodology, which has been described in the literature on many occasions as previously pointed out.

More specifically, the OMB has been validated in the laboratory against test bench instrumentation, in the field against altimeter satellite data, and in the field against commercial buoys \cite{rabault_openmetbuoy-v2021_2022}. This has confirmed the ability of the OMB to measure waves down to typically 0.5 cm amplitude at a 16 s period, as mentioned above (see \cite{rabault_openmetbuoy-v2021_2022}, and Fig. \ref{fig:wave_accuracy} reproduced from there).


\section*{Usage Notes}

GPS position data can be used without further considerations, as the GPS module itself performs all the data quality assurance necessary, and the position accuracy is more than enough to resolve trajectories given the 30 minute sample rate.

Similarly to other acceleration-based wave measurements, and in good agreement with the motion sensor datasheet, the OMB exhibits an acceleration noise spectral density that is constant (i.e., independent of frequency). As a consequence, the noise level for the wave elevation spectral density, which is obtained by integrating elevation twice in time, shows increased noise level in the low frequency range. This is a well-known effect \cite{rabault_openmetbuoy-v2021_2022,waseda_arctic_2017,nose_predictability_2018}. In order to mitigate this effect, we transmit wave vertical acceleration spectra over iridium, so that the raw acceleration spectra are available to the end user. Elevation spectra are computed from the acceleration spectra using a Python script that is included in the OMB GitHub repository. This makes it easy for the end user to observe the unfiltered noise level present in the wave spectra, before any further processing is applied. Our data processing code also generates a low frequency cutoff estimate to take into account this low frequency noise amplification. This cutoff is obtained as the first local minimum, if such a minimum exists, observed between the lowest-frequency bin and the peak of the wave elevation spectrum, as previously described in \cite{waseda_arctic_2017}. This allows us to separate between actual signal and noise. We underline that the unfiltered acceleration power spectral density data are always available to the user, so that a visual inspection of each individual spectrum before any filtering can be performed if necessary. Moreover, the code used for generating this low-frequency cutoff is part of the open source code we provide, so it can be inspected by the user.

As described in previous work \cite{rabault_dataset_2023}, the wave spectrum can be used both to measure waves themselves, and as a robust method to determine whether a buoy is still on an ice floe in the MIZ, or floating freely in open water after ice melting or breakup. Indeed, the sea ice acts as a lowpass filter and high frequency waves are very efficiently damped by the outer MIZ and filtered out by the ice floe response. As a consequence, a buoy on sea ice in the MIZ typically displays a spectrum that presents very little high-frequency energy content. By contrast, a buoy floating freely typically displays a significant amount of wiggling motion owing to its small size, and a significant high-frequency energy tail is then observed. As a consequence, the absence, or presence, of a high-frequency spectral tail is a robust indication of whether a buoy is on the ice or in the water, as illustrated in Fig. \ref{fig:overview_spectra}.

%

\section*{Code availability}

The OpenMetBuoy-v2021 used to collect all the data released here is fully open source. The hardware, firmware, and data post-processing details are available in the public GitHub repository of the project: \url{https://github.com/jerabaul29/OpenMetBuoy-v2021a}. The OpenMetBuoy firmware can be compiled with the Arduino IDE v1.8.19 available at \url{https://www.arduino.cc/en/software} (see instructions on the OMB GitHub repository for library installations). The buoy firmware is written in C++, while post-processing tools and data decoders are written in Python3.

All the code used to format and prepare the data released in this paper is available on the public GitHub repository associated with this article: \url{https://github.com/jerabaul29/2024_OpenMetBuoy_data_release_MarginalIceZone_SeaIce_OpenOcean}. This code is written in Python3, and does heavy use of the Trajectory Analysis (TrajAn) software package \url{https://github.com/OpenDrift/trajan}.

Similarly to \cite{rabault_dataset_2023}, we will offer reasonable support regarding the data and its use through the Issues tracker of the data repository associated to this article at \url{https://github.com/jerabaul29/2024_OpenMetBuoy_data_release_MarginalIceZone_SeaIce_OpenOcean}. Moreover, we invite readers in need of specific help to contact us there. In addition, we plan on releasing extensions to this dataset periodically as more data are collected. We invite scientists who own similar data and are willing to release these as open data and open source materials to contact us so that they can get involved in the next data release we will perform, and we hope to be able to support and encourage further community-wide data releases in the future.

While there are already a significant number of datasets available about sea ice drift and waves in ice (for example, \cite{SIPEXII,long_term_meas_Thomson,wave_driven_flow_data,rabault_dataset_2023}), these are scattered across the internet, and often hard to find. Therefore, we maintain an index of similar data at \url{https://github.com/jerabaul29/meta_overview_sea_ice_available_data}. We invite readers aware of additional similar data to submit an issue or pull request on the corresponding GitHub repository, adding additional pointers to similar datasets.


\bibliography{sample}

\section*{Acknowledgements}

We thank the following projects and agencies, that partly funded this work:

\begin{itemize}
    \item The Research Council of Norway, project 276730 (The Nansen Legacy), 237906 (CIRFA), 303411 (Machine Ocean), 301450 (FOCUS),
    \item the Australian Antarctic Science Program under Project AAS4593,
    \item YK was supported by Lundin Energy Norway (grant 802144).
    \item Some of this work, in particular TN work, was a part of the Arctic Challenge for Sustainability II (ArCS II) Project (Program Grant Number JPMXD1420318865). A part of this study was also conducted under JSPS KAKENHI Grant Numbers JP 19H00801, 19H05512, 21K14357, and 22H00241. A part of this study was conducted under ESA ARKTALAS Hoavva project number: 4000127401/19/NL/LF. 
\end{itemize}

We thank the crews of the R/V Kronprins Haakon, the R/V Polarstern, the R/V Oden, and the R/V Helmer Hansen for their invaluable help during fieldwork and instrument deployment.


\section*{Author contributions statement}

JR was the main initiator of the manuscript and designed the data collection. JR put the data in the common format, prepared the technical and data management aspects. JR designed the OMB-v2021 (with help from GH, TN, OG and MM). JR, CT, MI, GH, TN, HTB, TKo, SO, JV, MM built specific instruments used in deployments. JR, CT, MI, YK, AJ, ØB, MH, AK, MJ, TE, KDF, JRo, LE, TW, AB, JV, JWP, MM designed, obtained funding for, organized, and lead field campaigns that deployed instruments. JR, CT, MH, TN, YK, ØB, HTB, MH, DD, MJ, KFD, MDM, ESUR, JL, VCMdA, TKa, JWP, and MM deployed instruments. JR, CT, GH, TN, DD, TD, JV, JWP, MM, collected and gathered data from individual deployments reported in the manuscript. JR wrote the manuscript, with help from ØB, and feedback from all co-authors on the deployments each of them had participated into. All authors reviewed the manuscript and participated in iterating the manuscript.


\section*{Competing interests}

The authors declare no competing interests. In particular, the authors have no commercial involvement in the LabMaker OMB commercial product.


\newpage

\section*{Figures \& Tables}

\begin{table}[ht]
\centering
\begin{tabular}{|l|l|}
\hline
\hline
size (default box) & 10 $\times$ 10 $\times$ 12 cm \\
\hline
weight (battery and box dependent) & 0.5-1.0 kg \\
\hline
\hline
microcontroller & Ambiq Apollo 3 \\
\hline
communications & Iridium SBD \\
\hline
GNSS chip & ZOE-M8Q \\
\hline
motion sensor chip & ISM330DHCX \\
\hline
\hline
battery (standard) & 2-3 LSH20 \\
\hline
autonomy (standard) & 7.5+ months \\
\hline
\hline
GNSS sample rate (standard) & once every 15 minutes \\
\hline
wave 1D spectrum sample rate (standard) & once every 2 hours \\
\hline
\hline
hardware cost & around 650 USD \\
\hline
total cost of ownership, 3 months deployment & around 1000 USD \\
\hline
\hline
motion sensor raw data rate & 800 Hz \\
\hline
Kalman filter refresh rate & 100 Hz \\
\hline
wave acceleration sample rate & 10 Hz \\
\hline
wave spectrum data sample duration & 20.48 minutes \\
\hline
\hline
online open source project repository & \url{https://github.com/jerabaul29/OpenMetBuoy-v2021a} \\
\hline
\hline
\end{tabular}
\caption{\label{tab:omb_properties}Key properties of the OMB-v2021.}
\end{table}

\begin{table}[ht]
\centering
\begin{tabular}{|l|p{12.0cm}|}
\hline
link & URL \\
\hline
\hline
GitHub: OMB main page & \url{https://github.com/jerabaul29/OpenMetBuoy-v2021a} \\
\hline
commercial OMB & \url{https://www.labmaker.org/products/openmetbuoy} \\
\hline
GitHub: data release & \url{https://github.com/jerabaul29/2024_OpenMetBuoy_data_release_MarginalIceZone_SeaIce_OpenOcean} \\
\hline
GitHub: MIZ data overview & \url{https://github.com/jerabaul29/meta_overview_sea_ice_available_data} \\
\hline
GitHub: TrajectoryAnalysis & \url{https://github.com/OpenDrift/trajan} \\
\hline
data release archive & \url{https://adc.met.no/datasets/10.21343/w2se-b681} \\
\hline
\hline
\end{tabular}
\caption{\label{tab:urls}URLs for webpages and code release. Note that individual URLs for each dataset hosted on the Arctic Data Center are gathered in Table \ref{tab:summary}.}
\end{table}

\begin{table}[ht]
\centering
\begin{tabular}{|l|l|l|l|l|}
\hline
deployment & doi link & start & buoys on ice & buoys in water \\
\hline
\hline
CIRFA\_UIT & \url{https://adc.met.no/datasets/10.21343/x2c1-rb34} & 2022-04 & 19 & 0 \\
CIRFA\_METNO & \url{https://adc.met.no/datasets/10.21343/v2ca-3h77} & 2022-04 & 0 & 15 \\
AWI\_UTOKYO & \url{https://adc.met.no/datasets/10.21343/zt89-k846} & 2022-07 & 15 & 0 \\
UIO\_DOFI & \url{https://adc.met.no/datasets/10.21343/8nsc-vg26} & 2022-08 & 14 & 0 \\
CAGE & \url{https://adc.met.no/datasets/10.21343/47ky-m450} & 2022-08 & 0 & 6 \\
UIB\_METNO & \url{https://adc.met.no/datasets/10.21343/ndmw-4z34} & 2022-10 & 0 & 2 \\
CHALMERS & \url{https://adc.met.no/datasets/10.21343/qtmp-jp49} & 2023-05 & 7 & 0 \\
AWI\_UOM & \url{https://adc.met.no/datasets/10.21343/sdxx-c192} & 2023-08 & 1 & 0 \\
\hline
\hline
\multicolumn{3}{|l|}{total: 79 buoys} & 56 & 23 \\
\hline
\end{tabular}
\caption{\label{tab:summary}Summary of the deployments included in this data release.}
\end{table}






\end{document}